\documentclass[aps,twocolumn,floatfix,showpacs]{revtex4}
\usepackage{graphicx,bm}
\usepackage{amsmath}
\usepackage{amssymb}
\begin{document}
\title {Stability of the two-dimensional Bose gases in the resonant regime} 
\author{Ludovic Pricoupenko$^{1}$ and Maxim Olshanii$^{2}$}
\affiliation
{$^{1}$Laboratoire   de   Physique  Th\'{e}orique de    la Mati\`{e}re
Condens\'{e}e, Universit\'{e} Pierre et Marie Curie, case courier 121,
4 place  Jussieu, 75252 Paris Cedex  05, France.\\ $^{2}$Department of
Physics \& Astronomy,  University of Southern  California, Los Angeles,
CA 90089-0484, USA}
\date{\today}
\begin{abstract}
We consider a two-dimensional Bose gas formed in a planar 
atomic trap in conditions where the two-dimensional scattering 
length exceeds all other microscopic length scales in the system
and, accordingly, the gas parameter assumes relatively high values. 
We show that, unlike in the three-dimensional case, for sufficiently 
low areal densities the two-dimensional gas remains stable against 
collapse even in the resonant regime. Furthermore, we evaluate the 
three-body recombination rate that sets the upper limit for the 
life-time of two-dimensional resonant atomic condensate. 
\end{abstract}

\maketitle

\section{Introduction}

Since the achievement  of Bose Einstein  Condensation (BEC) of neutral
atoms  in  1995, the  field   of  cold  atoms has  been  substantially
developed, both in the parameter range  (number of atoms, temperature,
{\it etc}), and   in the variety  of systems  studied.  In particular,
low-dimensional atomic  systems   became  a subject  of   significant
interest as a consequence of the existence of  nontrivial strong phase 
fluctuations and also due to the peculiar scattering properties
\cite{QGLD2003}. Two-dimensional (2D) Bose gases play a special role in 
this       sub-field   thanks       to   a    rich      phase  diagram
\cite{Svistunov}.  Experimentaly, a two-dimensional quasi-condensate has
been first  achieved with spin-polarized hydrogen  atoms adsorbed on a
liquid  $^4$He film  \cite{Safonov}.  Several groups  have now reached
the 2D regime with ultra cold atoms trapped using evanescent light
waves \cite{Grimm_2D} or optical lattices \cite{Jean1}. In the latter 
trap, thermaly activated  vortices  which plays  a   central role  in  
the Berezinsky-Kosterlitz-Thouless  phase transition  have been  
observed  experimentaly \cite{Jean2}. Ultra-cold atoms in two  
dimensions are also interesting  due to the fact that the atom-atom 
effective  coupling constant can be easily tuned \cite{Petrov1,Petrov2} 
by varying the axial confinement of the     trap      and/or   by  
using    a    Feshbach  resonance \cite{Cornish}. Consequently, 
ultra-cold  atoms  experiments  open interesting perspectives  for  
a  detailed  study  of  atomic BEC in strictly   ({\it i.e.}  with    
no excited  transverse  modes present) two-dimensional  configurations  
in  a  broad  range of   experimental parameters.

Using the 2D equation of state, we show that  in the resonant regime 
where the two-dimensional scattering length exceeds the size of the 
transverse confinement (and becomes the only relevant length scale 
associated with the interatomic interactions) the gas remains stable 
(at least for sufficiently small areal densities), even 
though the corresponding three-dimensional scattering length becomes 
negative in this regime. Moreover, since in the resonant regime the 
2D gas parameter becomes relatively high, the formation of dimers in 
three-body collisions becomes relevant \cite{Giorgini}. Therefore, we 
study further the stability of the resonant two-dimensional condensate 
{\it vis a vis} this process.

The paper is composed of two parts. In the  first part, we review 
common low energy collisional features for two  ultra-cold atoms in a
planar  harmonic  guide and  give the definition  of the  2D scattering
length  ($a_{2D}$).  The  effective  interaction  between two confined
atoms    is described using the zero-range  approximation  within the
$\Lambda$-potential  formalism.   The    resonant regime  for  two-body
scattering is  defined by the condition $a_{2D} \gg  a_z$. A bound state
of  vanishing  energy and  large  spatial extension (of the  order of
$a_{2D}$) emerges under these conditions:  this is nothing else but a 
dimer which is populated by  three body processes in  a dilute resonant gas. 

In the second part, we  consider the 2D  atomic  Bose gas of areal density
$n$.  For low areal density systems where the interparticle distance is much
larger than the transverse confinement $na_z^2\ll 1$, the interaction can be
described in terms of the 2D scattering length only. For small gas parameters  
$n  a_{2D}^2 \ll 1$, the   two-dimensional equation of state shows  no sign 
of  a collapse  even  though the corresponding three-dimensional scattering 
length $a_{3D}$    is negative in the    two-dimensional resonant regime.
Finally,    we  evaluate the   rate   of   three-body  atomic recombinaison 
into the shalow dimer and give an  estimate for the life time of a 
two-dimensional  atomic condensate at the onset of the resonant regime.

\section{Two-body scattering in a two-dimensional planar wave-guide}
\subsection{Planar wave-guide}

We consider two atoms in an harmonic planar waveguide characterized by
the  angular atomic frequency  $\omega_z$  in the transverse direction $z$.
The characteristic   length   is given by   the size    of the  ground
transverse mode  $a_z= \sqrt{\displaystyle  \frac{\hbar}{m\omega_z}}$.  
The two atoms have the same mass $m$ and are located  at positions ${\bf R}_1$
and ${\bf   R}_2$,  respectively.  We further  decompose  the position
vectors into  a   sum of longitudinal   (parallel to  $xy$-plane)  and
transverse (along $z$ axis) components: ${\bf R}_1 = {\bf r}_1 + z_1
\hat{\bf e}_z$ and ${\bf R}_2 = {\bf r}_2 + z_2 \hat{\bf  e}_z$,  where  
$\hat{\bf e}_z$ is  a unit vector  in the transverse direction.   This
system can be considered as 2D when the energy  of the relative motion
$E$ is close to the threshold between the monomode and non-propagating
regimes: $\epsilon \equiv E-\hbar\omega_{z}/2 \ll\hbar\omega_{z}$.  Then, for $|{\bf r}_1-{\bf r}_2| \gg
a_{z}$, the two-body wave function factorizes as:
\begin{equation}
\Psi_2({\bf R}_1,{\bf R}_2) = \phi_0(z_1) \phi_0(z_2) \psi_2({\bf r}_1,{\bf r}_2) \, ,
\label{eq:psi3D}
\end{equation}

\subsection{Two-dimensional scattering amplitude and two-dimensional 
scattering length}

In a scattering process at energy $\epsilon = \hbar^2 k^2/m$  
(with $k a_z \ll 1$), and for large relative distance ($r \gg  a_z$), 
the 2D wave function is \cite{Petrov1,Petrov2,LOWD}:
\begin{equation}
\psi_2({\bf r}) = \exp(i{\bf k}.{\bf r}) + \frac{i \pi}{2 \ln(-i k q a_{2D})} H_0^{(1)}(kr)  \, .
\label{eq:scatt_ampl}
\end{equation} 
In Eq.(\ref{eq:scatt_ampl}) ${\bf   r}={\bf  r}_1-{\bf r}_2$  are  the
relative coordinates,  the  constant $a_{2D}$ is  the  ``2D scattering
length'' and $q = e^{\gamma}/2$, where $\gamma$ is the Euler constant. It can be
shown \cite{Petrov2,LOWD} that $a_{2D}$ is a function of the characteristic
length of the trap $a_z$ and also of  the three dimensional scattering
length $a_{3D}$ with \cite{2Dconstant}:
\begin{equation} 
a_{2D}  =  2.092 \, a_z \exp\left(-\sqrt{\frac{\pi}{2}} 
\frac{a_z}{a_{3D}}\right) \, . \label{eq:a2D}
\end{equation}

\subsection{Zero-range approximation in two dimensions}

Let  us   introduce  the  notion    of a   zero-range  two-dimensional
interaction potential for the two trapped  atoms. For this purpose, we
use the  fact that the radius of   the interatomic forces  $R$ is very
small as compared to all the lengths considered in this paper, so that
the  3D   wave function is   a solution   of the  free Schr\"{o}dinger
equation ($r \gg R$):
\begin{equation}
-\frac{\hbar^2}{2 m} (\Delta_1+\Delta_2) \Psi_2 
+ \frac{1}{2} m \omega_z^2 (z_1^2+z_2^2) \Psi_2 = E \Psi_2 \, . 
\end{equation}
Consequently,  for  $r \gg a_z$ and $\epsilon \ll \hbar \omega_z$,   the  2D wave function
$\psi_2$ is also solution  of the free  Schr\"{o}dinger equation.  Hence,
$\psi_2$ has a logarithmic behavior for small interparticle distances (in
the scattering process this  corresponds to $k  r \ll 1$ but $r \gg a_z$).
The idea of the  zero-range approach is to  consider the  extension of
$\psi_2$ defined  as a solution  of the  free Schr\"{o}dinger equation $\forall
r>0$ and to impose the correct behavior of the  wave function for $r \gg
a_z$  from  a contact condition   taken  formally at  $r=0$. Using the
expression   of the  scattering  states  in   Eq.(\ref{eq:scatt_ampl})
extended in the region $kr \ll 1$, one  can verify that all the two-body
states which are  linear combinations of (\ref{eq:scatt_ampl})  can be
deduced from the contact condition:
\begin{equation}
\psi_2({\bf r}_1,{\bf r}_2) = A_2 \ln \left( \frac{r}{a_{2D}} \right) + {\cal O }(r) \, ,  
\label{eq:contact1}
\end{equation}
where $A_2$ depends on the state considered and also can be a function
of the   center of mass  coordinates.  The  zero range two-dimensional
potential introduced in Ref.\cite{lambdapot} allows to incorporate the
contact condition  (\ref{eq:contact1}) in the Schr\"{o}dinger equation
defined $\forall r \geq 0$ (at $r=0$, the identity $\Delta \ln(r) = 2 \pi \delta^{(2)}({\bf
r})$ is used). It has the following expression:
\begin{equation}
V_{12}^{\Lambda}({\bf r}\,) = - \frac{2\pi\hbar^2}{m} \, \frac{\delta^{(2)}({\bf r}\,)}{\log(\Lambda a_{2D})} 
\, \left[ 1 - \log(\Lambda r) r \frac{\partial}{\partial r}  \right] \, ,
\label{eq:2D_ppot}
\end{equation}
where $\Lambda$  is   an arbitrary constant.   This  potential acts  on wave
functions of the form (\ref{eq:contact1}) as:
\begin{equation}
\langle {\bf r}_1,{\bf r}_2 | V_{12}^{\Lambda} | \psi_2 \rangle = \frac{2 \pi \hbar^2}{m} A_2 \, \delta^{(2)}({\bf r}) \, ,
\end{equation}
and  gives a  result independent of  $\Lambda$. Let  us emphasize that  this
property is not general: the action of $V_{12}^{\Lambda}$ on a two-body wave
function which has not the singularity given by Eq.(\ref{eq:contact1})
leads to a $\Lambda$-dependent result, showing also  that such wave function
is not in the proper Hilbert space.

\subsection{Resonant regime: shallow bound state}

In the  regime  of large  2D  scattering  length $a_{2D} \gg a_z$  which
corresponds to    the situation  of    negative 3D  scattering  length
$a_{3D}<0$, the scattering cross section  is maximum for $k = \kappa_B \equiv (q
a_{2D})^{-1}$.  In   this resonant regime, there   exists a low energy
two-body  bound  state (refered hereafter as  the  dimer state) with a
spatial extension of the order  of $a_{2D}$. Its wave function $(\phi_B)$
and binding energy $(\epsilon_B)$  can be deduced  from the contact condition
(\ref{eq:contact1})     or     also  from       the   pseudo-potential
(\ref{eq:2D_ppot}). One obtains:
\begin{equation}
\phi_B(r_{12}) = \frac{\kappa_B}{\sqrt{\pi}} {\rm K}_0\left(\kappa_B r_{12} \right) , \quad \mbox{and} \quad \epsilon_B= -\frac{\hbar^2\kappa_B^2}{m} \, .
\label{eq:dimer}
\end{equation}
Now,  the key  point for  the  following part   is  that the state  in
Eq.(\ref{eq:dimer}) has  an extension of  the order of  $a_{2D}$: as a
consequence it corresponds to the asymptotic part of a physical one in
the regime:
\begin{equation}
a_{2D} \gg a_{z} \, . 
\label{eq:univ}
\end{equation}
As a consequence of the exponential behavior in Eq.(\ref{eq:a2D}), this
resonant regime is achieved  for negative three dimensional scattering
length and:
\begin{equation}
0 < - a_{3D} < a_{z} .
\label{eq:resonant}
\end{equation}

\section{Is the dilute resonant two-dimensional Bose gas stable?}  

\subsection{Equation of state: stability against collapse}

We consider an atomic Bose gas characterized by  an areal density $n$
and confined in the planar wave guide with $n a_z^2 \ll 1$, so that only
the ground transverse  mode is populated.  In the dilute limit defined
by a small 2D gas parameter:
\begin{equation}
n a_{2D}^2 \ll 1 \, ,
\label{eq:dilute}
\end{equation}
the 2D equation of state at zero temperature is \cite{Popov,Mora,2D}:
\begin{equation}
n = \frac{m\mu}{4\pi \hbar^2}\ln\left(\frac{4 \hbar^2}{e^{(2\gamma+1)}m\mu a_{2D}^2} \right) \, .
\label{eq:popov}
\end{equation}
For sufficiently   small areal densities  such that Eq.(\ref{eq:dilute})  is
satisfied,  Eq.(\ref{eq:popov})  shows  that  $m\mu a_{2D}^2/\hbar^2 \ll1$  and
$\partial n/\partial\mu>0$, $\forall a_{2D}$. Consequently for a vanishing areal density, the
system is thermodynamically stable which means an absence of collapse. 
This  is in  particular  true  in  the  regime  of negative three
dimensional scattering length $a_{3D}<0$,  where an instability occurs
for the homogeneous gas in absence of trapping. Due to the exponential
behavior of $a_{2D}$ with  respect  to $a_{3D}$ in  Eq.(\ref{eq:a2D}),
the  condition (\ref{eq:dilute}) is too   stringent to reach for  $0<-
a_{3D} \ll  a_z$ and the  more interesting situation  corresponds to the
onset of the resonant regime:
\begin{equation}
0 < - a_{3D} \lesssim a_z \, .
\label{eq:onset}
\end{equation}
However one  can  notice that  even  at the  onset,  the  condition 
defined by Eq.(\ref{eq:dilute}) is severe and is difficult to satisfy 
for clouds involving thousand of atoms (as a consequence of the radial 
trapping) \cite{trap}.

However, even  if the  2D  Bose homogeneous   gas is  stable   against
collapse        in     the      peculiar    regime      defined     by
Eqs.(\ref{eq:dilute},\ref{eq:onset}),  its  stability   is limited  by
three body   recombinaison    processes  into   the    shallow   dimer
Eq.(\ref{eq:dimer}).  In the next  section we adress this question and
give an estimation for the life-time in this regime.  Hence, we give a
limit on the possible occurence of an atomic condensate for increasing
values of the  2D  gas parameter obtained by  a  variation of   the 2D
scattering length.

\subsection{Three-body recombination rate: stability against formation 
of dimers}

The problem  that we  are   considering is reminescent  of the   three
dimensionnal resonant   Bose   gas where three    body  recombinations
populate a  dimer  state  of  vanishing  energy   and diverging  width
\cite{Fedichev}. More precisely, the transverse length $a_z$ plays the
role of an effective range and  Eq.(\ref{eq:univ}) is analogous to the
resonant regime in 3D  where $a_{3D}\gg R$.   However the two situations
are  not    completely similar. Indeed,  in   the   2D resonant regime
Eq.(\ref{eq:univ})   a single  parameter $a_{2D}$  is sufficient for a 
characterization of the low energy properties in few- and many-body  
systems \cite{Adhikari}, while in 3D  at least one more
parameter is  needed \cite{Bedaque}. Moreover,  in  2D the  low energy
two-body  cross section strongly  depends on  the energy  so that  one
expects a strong variation of  the recombination rate with respect  to
the gas   parameter which  fixes the  typical   collisional energy.  A
straighforward dimensional analysis  gives  the following law for  the
recombination constant:
\begin{equation}
        \alpha_{rec} = F(n a_{2D}^2) \frac{\hbar}{m} a_{2D}^2 \, ,
\label{eq:thelaw}
\end{equation} 
and  the purpose of the subsequent  three-body analysis is to evaluate
the function $F$.

We consider three  atoms $(1,2,3)$  of  equal mass $m$  which after  a
collision lead to  the formation of a dimer \cite{Clarification}. There   
are   three outgoing channel    for this process obtained by cyclic 
permutation from:
\begin{equation}
1 + 2 + 3 \longrightarrow 1 + (23)  \quad ({\mathcal C}) \, ,
\end{equation}
where $(23)$ denotes the  dimer    and $({\mathcal C})$  labels   this
particular  channel.   The problem  can   be solved using  the Faddeev
equations \cite{Faddeev}  in  spatial coordinates.  This   approach is
very efficient for the determination in the configuration space of the
two universal three-body bound states in two dimensions \cite{Nielsen}
(these states where first obtained in the momentum representation in
Ref.\cite{Adhikari_1}).  The  positions  of  the atoms  $(1,2,3)$  are
defined  on  the plane by  the  coordinates $({\bf r}_1,{\bf r}_2,{\bf
r}_3)$   respectively.  The  wave  function factorizes   in a $z$-part
frozen  in  the ground state   of the  transverse  harmonic oscillator
(omitted in the following) and a 2D part $\Psi_3({\bf r}_1,{\bf r}_2,{\bf
r}_3)$. The wave function $\Psi_3$ is an eigenstate of the Hamiltonian:
\begin{equation}
        H_3 =  - \sum_{i=1}^3 \frac{\hbar^2}{2 m} \Delta_{{\bf r}_i} + 
V_{12}^{\Lambda} + V_{13}^{\Lambda} + V_{23}^{\Lambda}  .
\label{eq:H3}
\end{equation}
To  be consistent with the 2D  regime, the  eigen-energy of this state
must be below the transverse level spacing, $\epsilon \ll \hbar \omega_z$. The potential
part      in  Eq.(\ref{eq:H3})   imposes       the  contact  condition
(\ref{eq:contact1}) for cyclic permutations of the three particles. In
the center of mass  frame, the wave function, which  is symmetric as a
consequence of the symmetrization postulate is decomposed as:
\begin{equation}
\Psi_3(1,2,3) = \psi({\bf u}_1,{\bf u}_2) + \psi({\bf u}_1',{\bf u}_2') + \psi({\bf u}_1'',{\bf u}_2'') \, , 
\end{equation}
with the Jacobi variables defined by:
\begin{equation}
{\bf u}_1 = \sqrt{\frac{2}{3}} (
{\bf r}_1-\frac{{\bf r}_2+{\bf r}_3}{2} )
\quad ; \quad {\bf u}_2 = \frac{1}{\sqrt{2}} ({\bf r}_2 - {\bf r}_3) .
\end{equation}
The other variables $({\bf  u}_1',{\bf u}_2')$ and $({\bf  u}_1'',{\bf
u}_2'')$  can be also introduced  via cyclic permutations of the three
atoms.  We  use  then the   hyperspherical  coordinates $\{\rho,\alpha,\alpha',\alpha''\}$
defined by:
\begin{equation}
        u_1 = \rho \cos(\alpha) \quad ; \quad u_2 = \rho \sin(\alpha) \, , 
\end{equation}
and hyperangles $\alpha'$ and $\alpha''$ are obtained by cyclic permutations. As
we are interested in a low energy process, the 3-body wave function is
a $s$-state which does not depend on the planar angles. Consequently, 
the kinetic operator in Eq.(\ref{eq:H3}) is simply
\begin{eqnarray}
&&\hat{T} = \frac{\hbar^2}{2 m} \left( - \frac{1}{\rho^{3/2}}
\partial_\rho^2( \rho^{3/2} . ) + \frac{\displaystyle \frac{3}{4} + \hat {\mathcal A}^2}{\rho^2} \right) ,\\
&&\mbox{with,} \quad \hat {\mathcal A}^2 = - \partial_{\alpha}^2 - \frac{2}{\tan(2 \alpha)} \partial_{\alpha} \, .
\end{eqnarray}
In this coordinate system, the contact condition reads
\begin{equation}
\Psi_3 = A_3(\rho) \ln\left( \frac{\alpha \rho \sqrt{2} }{a_{2D}} \right) + {\cal O }(\alpha) \, ,
\label{eq:contact2}
\end{equation}
and the contact conditions for $\alpha'$  and $\alpha''$ have an analogous form.
We are going to show now, that the  knowledge of the function $A_3(\rho)$
is sufficient for  the  determination of the three  body recombination
rate.   For  this purpose,  we   use  the expression  of  the on-shell
$t$-matrix element from   the initial channel of  three  atoms  in the
continuum of    positive total energy  $\epsilon$,  to  the  outgoing channel
$(\mathcal C)$
\begin{equation}
        t_{\mathcal C} = \langle \Psi_{(\mathcal C)}^{\rm out} | V_{(\mathcal C)} | \Psi_3 \rangle \, .
\label{eq:tif}
\end{equation}
In Eq.(\ref{eq:tif}), $\Psi_3$ is the exact  3-body   eigen-function of the Hamiltonian
(\ref{eq:H3}) corresponding to the initial state, and  $\Psi_{(\mathcal C)}^{\rm out}$ is 
the  non-interacting  3-body wave function for the  outgoing channel $(\mathcal C)$:
\begin{equation}
\langle{\bf u}_1,{\bf u}_2|\Psi_{(\mathcal C)}^{\rm out} \rangle  =  \frac{1}{\sqrt{3}}
\exp \left( \displaystyle i {\bf k}.{\bf u}_1 \sqrt{\frac{3}{2}} \right)  
\phi_B(|{\bf r}_2-{\bf r}_3|) \, ,
\label{eq:outgoing}
\end{equation}
where the wave number $k$ is  obtained from the energy conservation in
the   3-body process:  $k   = \displaystyle\frac{2\kappa_B}{\sqrt{3}}$. 
The interacting potential $(V_{(\mathcal C)})$ which appears in Eq.(\ref{eq:tif}) is the scattering 
potential for channel $({\mathcal C})$ \cite{Taylor}:
\begin{equation}
V_{(\mathcal C)} =  V_{12}^{\Lambda} + V_{13}^{\Lambda} \, .
\end{equation}
Summing up the contributions of the three possible outgoing channels, 
one obtains the total $t$-matrix element:
\begin{equation}
t_{if} = 6  \langle \Psi_{(\mathcal C)}^{\rm out} | V_{13}^\Lambda | \Psi_3 \rangle \, .
\label{eq:tif1}
\end{equation}
From Eqs.(\ref{eq:contact2},\ref{eq:tif1}), the total $t$-matrix element 
can be deduced from:
\begin{equation}
t_{if} = \frac{4 \pi^2 \sqrt{3} \hbar^2}{m} \int_0^\infty\!\!\!  d\rho \, \rho J_0(\frac{\kappa_B \rho}{\sqrt{2}})   
\phi_B(\rho\sqrt{\frac{3}{2}}) A_3(\rho) \, .
\label{eq:t_if}
\end{equation}
This   relation can be understood qualitatively   if  one remarks from
Eq.(\ref{eq:contact2}), that $A_3(\rho)$   is linked to  the probability  of
finding one atom  at   distance $\rho$ from    a colliding pair  with  an
interparticle spacing smaller than $a_{2D}$.  Then, as the integration
of this function  in (\ref{eq:t_if}) is over  lengths of the  order of
$a_{2D}$, the  value of $t_{if}$ depends  crucially on the probability
of finding three atoms at relative  distance $a_{2D}$, which coincides
also with  the  extension  of    the  dimer.  Note  that  the   result
(\ref{eq:t_if}) does not depend on $\Lambda$  as expected for a theory which
provides the exact contact conditions on the wave function.

\begin{figure}

\centerline{\includegraphics[width=7cm]{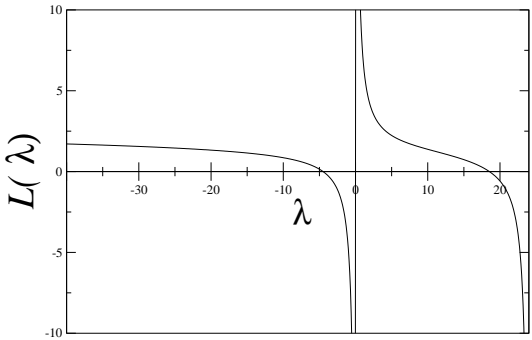}} 

\caption{The two lowest branches of the function 
$L$ -- left hand side of Eq.(\ref{eq:eigenvalue}).}
\label{fig:Lambda}
\end{figure}

\begin{figure}

\centerline{\includegraphics[width=8cm]{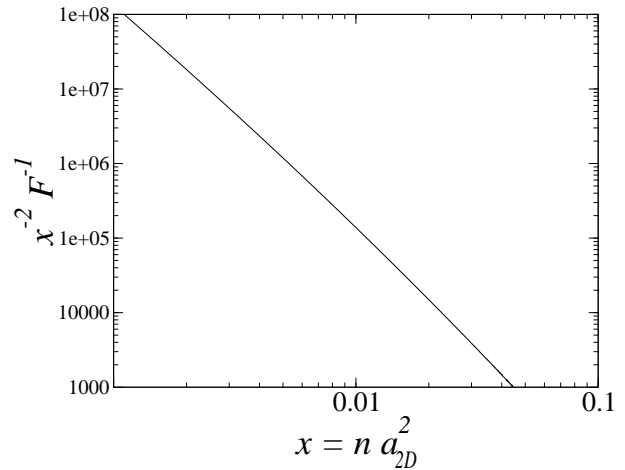}}

\caption{Function $1/(x^2F(x))$ entering in Eq.(\ref{eq:tau})}
\label{fig:universal}
\end{figure}

The determination  of the function $A_3$ can  be in principle obtained
from  the decomposition of  the  wave function over a hyperspherical
harmonics   basis  \cite{Nielsen}:
\begin{equation}
        \Psi_3 =  \sum_i \frac{f_i(\rho)}{\rho^{3/2}} \Phi_i(\rho,\alpha,\alpha',\alpha'')  \, ,
\label{eq:decomposition}
\end{equation}
where   $\Phi_i(\rho,\alpha,\alpha',\alpha'')=\phi_i(\rho,\alpha)+\phi_i(\rho,\alpha')+\phi_i(\rho,\alpha'')$ and for each value of the 
hyper-radius $\rho$, $\phi_i(\rho,\alpha)$ is an eigen-function of the operator  
$\hat{\mathcal A}^2$ with the eigen-value $\lambda_i(\rho)$  (this  step   is  
very  reminiscent  of  the     2-body problem). Taking into account the  
fact that the functions  $\phi_i(\rho,\alpha)$ are finite for $\displaystyle \alpha=\frac{\pi}{2}$, 
we obtain:
\begin{equation}
\phi_i(\rho,\alpha) = {\cal N}_i {\cal P}_{\nu_i}(-\cos(2\alpha)) \,,
\label{eq:phi}
\end{equation}
where ${\cal P}_{\nu_i}$ is the usual Legendre function and:
\begin{equation}
 \nu_i(\rho) = \frac{-1+\sqrt{1+\lambda_i(\rho)}}{2} \, .
\label{eq:nu}
\end{equation} 
For  a given value  of $\rho$, the normalization   constant ${\cal N}_i$ is
defined by integration over the angles $\alpha$ and $(\theta_1,\theta_2)$ (associated
to $({\bf u}_1,{\bf u}_2)$):
\begin{equation}
 \int_0^{\pi/2}\!\!\! d\alpha \int_0^{2\pi}\!\!\! d\theta_1 \int_0^{2\pi}\!\!\! d\theta_2 \,  | \Phi_i |^2 \sin(\alpha) \cos(\alpha) =
1 . \label{eq:norme}
\end{equation}
In the limit $\alpha \to 0$, the functions $\phi_i$ in Eq.(\ref{eq:phi}) exhibit a
logarithmic singularity:
\begin{eqnarray}
\phi_i(\rho,\alpha) = \frac{{\cal N}_i \sin(\pi\nu_i) }{\pi} &&
\bigg( \ln( \alpha^2 ) + 2 \gamma + 2 \psi(\nu_i+1) \nonumber \\ 
&& + \frac{\pi}{\tan(\pi \nu_i)} \bigg) + {\cal O }(\alpha) \, ,     
\label{eq:singular}
\end{eqnarray}
where $\displaystyle \psi(\nu)$ is  the logarithmic derivative of the gamma
function.   Hence, for  each  value of  $\rho$  the  contact condition in
Eq.(\ref{eq:contact2})  is satified  if  the  eigenvalue $\lambda_i(\rho)$   is a
solution of the equation:
\begin{eqnarray}
\gamma+\psi(\nu_i+1)+\frac{\pi}{2\tan(\pi \nu_i)} 
+\frac{\pi {\cal P}_{\nu_i}(1/2)}{\sin(\pi\nu_i)} = 
\ln\left(\frac{\rho \sqrt{2} }{a_{2D}} \right) .
\label{eq:eigenvalue} 
\end{eqnarray}
The left hand side of  Eq.(\ref{eq:eigenvalue}) is a function of $\lambda_i$
that  we  denote  by $L(\lambda_i)$.   This  function  has  several branches
associated to the harmonics in Eq.(\ref{eq:decomposition}) and the two
first ones are represented in Fig.(\ref{fig:Lambda}).
Solution  of  Eq.(\ref{eq:eigenvalue})   in  each   branch  defines  a
hyperspherical  harmonic $\Phi_i$. As the   initial non interacting state
corresponds  to three atoms in the  continuum with a vanishing energy,
the     most   important   weight    in      the decomposition      of
Eq.(\ref{eq:decomposition})  is  the  harmonic  corresponding to   the
lowest positive   eigenvalue.  We then  limit  our study   to a single
harmonic analysis with the wave function:
\begin{equation}
        \Psi_3 =  \frac{f(\rho)}{\rho^{3/2}} \Phi(\rho,\alpha,\alpha',\alpha'')  \, .
\label{eq:single}
\end{equation}
The approximation used here is linked to the fact that we can identify
a slow variable  ($\rho$)  and a  fast  variable ($\alpha$)  which are  almost
separable. The function $A_3(\rho)$  can then be deduced from: 
\begin{equation}
        A_3(\rho) = \frac{2 f(\rho)}{\pi\rho^{3/2}} \sin(\pi \nu) {\cal N}(\rho) \, . 
\end{equation}
For  each value  of   $\rho\geq0$,  we solve  Eq.(\ref{eq:eigenvalue}) in   the domain
$\lambda\in[0,24]$. This way  we construct the  function $\lambda(\rho)$ which leads
to an effective kinetic  barrier in the  one dimensional equation  for
$f(\rho)$~:
\begin{equation}
 - \partial_\rho^2f +\frac{\frac{3}{4}+\lambda(\rho)}{\rho^2} f = K^2 f \, .
\label{eq:F}
\end{equation}
In Eq.(\ref{eq:F}), $K$ is  the momentum defined by  the energy of the
process $\epsilon=\frac{\hbar^2K^2}{2m}$  ($K a_{2D}\ll 1$). In  the limit  of very
large hyper-radius, the  amplitude of $f$  is fixed by considering the
asymptotic behavior of the free incoming wavefunction, so that
\begin{equation}
        f(\rho) \operatornamewithlimits{\simeq}_{\rho \to \infty} 4 \sqrt{\frac{6\pi}{K^3}} \sin(K\rho+\theta) \, .
\label{eq:asympt}
\end{equation}
We apply  now this formalism by considering  three  atoms initially in
the 2D  atomic  Bose condensate of   areal density $n$ at  the  onset of the
resonant regime (\ref{eq:onset}). The  typical energy in the collision
process  is given by $\epsilon \simeq  3 \mu$. For simplicity the chemical potential
can be evaluated by use of the Schick's formula \cite{Schick} which is
an approximate form of the equation of state:
\begin{equation}
\mu \simeq  \frac{4 \pi \hbar^2 n}{\displaystyle m |\ln(n a_{2D}^2)|} \, .
\label{eq:Schick}
\end{equation}
The gas  parameter is small $n  a_{2D}^2 \ll 1$ which  implies also that
the collisional energy is  such that $\epsilon \ll|\epsilon_B| \ll \hbar \omega_{z}$. An analysis
similar to Ref.\cite{Fedichev} gives the number of processes per unit
time and surface and permits an identification  of the function $F(x)$
(where $x=na_{2D}^2$ is the 2D gas parameter)
entering   in the  recombination constant  $\alpha_{rec}$ (\ref{eq:thelaw})
with $F(x) = m^2  t^2_{if}/(3\hbar^2 a_{2D})^2$, and  the life-time of the
atomic Bose gas is~:
\begin{equation}
        \tau(x) \sim \frac{m a_{2D}^2}{3\hbar x^2 F(x)} \, .
\label{eq:tau}
\end{equation}
In Fig.(\ref{fig:universal}), we have plot the function
$1/(x^2F(x))$.

As an  example, we consider $^7$Li atoms  confined  in a wave-guide of
atomic frequency $\omega_{z} = 2\pi \times 250$~kHz ($a_z=76$~nm and is very much
larger than the range of the true interatomic forces $\sim 2.8$~nm) with a 2D gas parameter
$x=na_{2D}^2 = 5 \times 10^{2}$~cm$^{-2}$ and $a_{2D} = 10 a_z$. We obtain for
such configuration a very short life time $\tau \simeq 2 \times 10^{-2}$~s, showing
that the  2D  atomic  condensate decays rapidly  by creating dimers in three-body 2D 
recombination processes and not by experiencing a collapse phenomenon ($a_{3D}<0$).
For a smaller but also ``large'' 2D gas parameter as $x=10^{-2}$ with $a_{2D} = 10 a_z$, 
then   $\tau \simeq 3$~s showing that the atomic condenstae may  be  observable for relatively  high
values of the gas parameter. However, as  shown on Fig.(\ref{fig:universal}), $1/(x^2F)$ exhibits a
rapid decrease with respect to the gas  parameter $x$, so that for $x$
on the order of $0.1$, the atomic condensate decays very rapidly.

\section{CONCLUSION}

In  this paper we analyse   two  potential sources  of instability  of
two-dimensional  atomic  Bose condensates  in  the universal  resonant
regime of the large  two-dimensional scattering  length (corresponding
to  a negative three-dimensional scatterring  length  of the order  or
less than the  size  of  the  transverse confinement): a   macroscopic
collapse and  a loss of atoms  through formation of  shallow dimers in
three-body  collisions.  We show  that  (a) the gas is  stable against
collapse and that  (b) in  the above regime  the  upper limit  on  the
condensate  life-time  set  by the   three-body   recombination can be
relatively   long. Moreover, for a   trap  sufficiently deep  in  the
longitudinal  direction, dimers can be kept  in the trap  opening the
possibility  of a series of   recombination process with  
successive formations of three- and four-body bound states toward ``2D
Bosons droplets'' states such as predicted in Ref.\cite{Hammer}.

{\bf Acknowledgments}

L. P.  thanks  Bernard Bernu,  Yvan  Castin and  Gora Shlyapnikov  for
interesting discussions on the subject. {\it LPTL} is {\it UMR 7600 of
CNRS}. M.~O. was supported by the NSF grant {\it PHY-0070333}.


\begin{thebibliography}{99}

\bibitem{QGLD2003} D.S. Petrov, D.M. Gangardt and G.V. Shlyapnikov 2004 
Proceedings of the school ``Quantum Gases in Low Dimensions'',
(EDP Sciences: Les Ulis-France) J. Phys. IV  {\bf 116} 5; 
M.G. Moore, T. Bergeman and M. Olshanii {\it ibid} 67; Y. Castin {\it ibid} 87.

\bibitem{Svistunov} Nikolay Prokof'ev and Boris Svistunov, Phys. Rev. A, {\bf 66},
043608 (2002).

\bibitem{Safonov} A.I. Safonov, S.A. Vasilyev, I.S. Yasnikov,
I.I. Lukashevich and S. Jaakkola, Phys. Rev. Lett. {\bf 81}, 4545 (1998).

\bibitem{Grimm_2D} D. Rychtarik, B. Engeser, H.-C. N\"agerl, R. Grimm, Phys. Rev. 
Lett. {\bf 92}, 173003 (2004).

\bibitem{Jean1} S. Stock, Z. Hadzibabic, B. Battelier, M. Cheneau, and J. Dalibard, 
Phys. Rev. Lett. {\bf 95}, 190403 (2005). 

\bibitem{Jean2} Z. Hadzibabic, P. Kr\"uger, M.Cheneau, B. Battelier, J.B. Dalibard, Nature, {\bf 441}
1118 (2006).

\bibitem{Petrov1} D.S. Petrov, M. Holzmann and G.V. Shlyapnikov,
Phys. Rev. Lett., {\bf 84} 2551 (2000).

\bibitem{Petrov2} D. S. Petrov and G. V. Shlyapnikov, Phys. Rev. A 
{\bf 64}, 012706 (2001).

\bibitem{LOWD} L. Pricoupenko, {\sl in preparation}.

\bibitem{2Dconstant} The numerical  constant in Eq.(\ref{eq:a2D}) is related 
to the constant $B$ defined in Eq.(21) of Ref.\cite{Petrov2} by the expression:
$2.092=\sqrt{\pi/B}/q$, where $B=0.9049\hdots$ \cite{LOWD} (this value slightly 
differs from the numerical result $B=0.915$ of Ref.\cite{Petrov2}).

\bibitem{Cornish} S. L. Cornish, N. R. Claussen, J. L. Roberts,
E. A. Cornell, and C. E. Wieman, Phys. Rev. Lett. {\bf 85}, 1795 (2000).

\bibitem{Giorgini} S. Pilati, J. Boronat, J. Casulleras, and S. Giorgini,
Phys. Rev. A {\bf 71}, 023605 (2005). 

\bibitem{lambdapot} Maxim Olshanii and Ludovic Pricoupenko, Phys. Rev.
Lett.{\bf 88}, 010402 (2002).

\bibitem{Popov}  V. N. Popov in {\it Functional Integrals in Quantum
Field Theory and Statistical Physics} (D. Reidel Publishing,
Dordrecht, 1983).

\bibitem{Mora} C. Mora and Yvan Castin, Phys. Rev. A, {\bf 67}, 053615 (2003). 

\bibitem{2D} L. Pricoupenko, Phys. Rev. A {\bf 70}, 013601 (2004). 

\bibitem{trap} We consider an atomic condensate of $N$ atoms in a pancake 
trap of  angular atomic frequencies $\omega_z$  in  the axial direction and
$\omega_\parallel$  in the  radial  direction with  $\omega_z \gg \omega_\parallel$.  For a small areal
density at  the center  ($n a_z^2 \ll 1$)  ($n$  is  the atomic  density
integrated over the  $z$-direction at the center  of the trap),  using
Eq.(\ref{eq:Schick}) the local density approximation gives:
\begin{equation}
 N=\frac{4 \pi^2 n^2 a_{\parallel}^4}{|\ln(n a_{2D}^2)|} \, ,
\label{eq:LDA}
\end{equation}
where $a_\parallel   =  \sqrt{\hbar/(m \omega_\parallel)}$.   This  relation  imposes  a strong
constraint on the the axial frequency: for example with $N=2000$ and a
gas  parameter $x=na_{2D}^2=5 10^{-2}$,  with a longitudinal frequency
of $10$~Hz and $a_{2D} = 10 a_z$, we find an axial frequency $\omega_z \simeq 2\pi
\times  250~$kHz. This frequency  is   clearly   not reachable  in   actual
experiments  (see  for  example   Ref.\cite{Grimm_2D} where  the axial
frequency is of the order of $10$~kHz).

\bibitem{Fedichev} P.O. Fedichev, M.W. Reynolds, and G.V. Shlyapnikov,
Phys. Rev. Lett. {\bf 77}, 2921 (1996).

\bibitem{Adhikari} Sadhan K. Adhikari, T. Frederico and I.D. Goldman,
Phys. Rev. Lett. {\bf 74}, 487 (1995).

\bibitem{Bedaque} P.F.~Bedaque, E.~Braaten and H.W.~Hammer,
Phys. Rev. Lett. {\bf 85}, 908 (2000).

\bibitem{Clarification} For clarification, note that the present
analysis  does  not   hold for  Hydrogen  gas   adsorbed  on a  Helium
film. Indeed,  in  this situation,   the  range  of  the   interatomic
potential is of  the same order as  the transverse length and also  of
the 2D scattering length.

\bibitem{Faddeev} L.D.~Faddeev, Soviet Physics JETP {\bf 12}, 1014 (1961).

\bibitem{Nielsen} E. Nielsen, D. V. Fedorov and A.S. Jensen,
Phys. Rev. A {\bf 56}, 3287 (1997).
\bibitem{Adhikari_1}S. K. Adhikari, A. Delfino, T. Frederico, I. D. Goldman, 
and L. Tomio, Phys. Rev. A {\bf 37}, 3666 (1988).

\bibitem{Taylor} John R. Taylor in {\it Scattering Theory},John Wiley \&
Sons, Inc. (1972).

\bibitem{Schick} M. Schick, Phys. Rev. A {\bf 3}, 1067 (1971).

\bibitem{Brodsky} I.~V. Brodsky, M. Yu. Kagan, A.~V.~Klaptsov, R.~Combescot and X. Leyronas, Phys. Rev. A {\bf 73}, 032724 (2006).

\bibitem{Hammer} H. W. Hammer and D. T. Son, Phys. Rev. Lett. {\bf 93}, 250408 (2004).

\end{thebibliography}
\end{document}